\documentclass{WileyMSP-template}

\usepackage[]{subfigure}
\usepackage{xcolor}
\usepackage{hyperref}
\hypersetup{colorlinks,linkcolor={blue},citecolor={blue},urlcolor={red}} 

\begin{document}


\title{Tuning magnetic states in CrI$_3$ bilayers via Moir\'e patterns}

\maketitle


\author{A. M. Le\'on}
\author{E. A. Vel\'asquez* }
\author{F. Caro-Lopera }
\author{J. Mej\'ia-L\'opez}



\begin{affiliations}
Dr. A. M. Le\'on\\
Max Planck Institute for Chemical Physics of Solids, Dresden, Germany.\\

Prof. E. A. Vel\'asquez\\
Grupo MATBIOM, Facultad de Ciencias Básicas, Universidad de Medell\'in, Medell\'in, Colombia.\\
Email Address: evelasquez@udemedellin.edu.co\\
Prof. F. Caro-Lopera\\
Facultad de Ciencias B\'asicas, Universidad de Medell\'in, Medell\'in, Colombia.\\

Prof. J. Mej\'ia-L\'opez\\
Centro de Investigaci\'on en Nanotecnolog\'ia y Materiales Avanzados CIEN-UC, Facultad de F\'isica, Pontificia Universidad Cat\'olica de Chile. CEDENNA, Santiago, Chile.\\

\end{affiliations}

\justify

\begin{abstract}
Commensurable twisted bilayers can drastically change the magnetic properties of 
chromium trihalide layered compounds, which opens novel opportunities for tuning magnetic states through layer rotations. Here, we introduce a mathematical approach to obtain Moir\'e patterns in twisted hexagonal bilayers by performing a certain commensurable rotation $\theta$ over one layer. To test our approach, we apply it to find Moir\'e structures with $\theta=21.79^{\circ}$ and $32.20^{\circ}$ in the phases R$\bar{3}$ and C2/m of CrI$_{3}$. 
For comparison purposes, we also consider a non-shifted CrI$_{3}$ structure.
Electronic and magnetic properties of the so-obtained systems are computed by \textit{Ab-Initio} methodologies. Results show the presence of  rotation-angle-dependent magnetic configurations and steep modifications of the dispersion bands due to variations in the nearest and next nearest distances among layers of Cr atoms. Modifications obtained from these commensurable rotations are discussed on the basis of competition among different energy contributions due to changes in the 
atomic arrangements.

\end{abstract}


\section{Introduction}

Commensurable rotated layers of graphene have promoted relevant research during the last decades. Thus, the so-called twistronics became popular after the discovery of the magic angle at around $1.1^\circ$ for superconductivity of two projectively coincident rotated layers of graphene 
\cite{Cao2018,Yankowitz2019}.
Other interesting phenomena such as ferromagnetism \cite{gonzalez2017electrically,Sharpe2019}, spin textures \cite {yananose2021chirality}, flat bands \cite{lucignano2019crucial,Haddadi2020}, van Hove singularities \cite{jones2020observation, xu2021tunable}, and correlated insulating phases \cite{Cao2018a,Xie2020} have also been found in this system. Using the same commensurable hexagonal cell via atomic networks, a number of theoretical and experimental works have been reported in other crystallographic structures. For example, in hexagonal boron nitride structures, the presence of moir\'e patterns leads to the formation of electronic minibands \cite{Mishchenko2014}; and systems such a twisted bilayer MoS$_2$ show changes in their phonon properties due to the periodic potentials of the moir\'e patterns \cite{Lin2018}. Experimental evidence of interlayer valley excitons trapped in a moiré potential has also been found in MoSe$_2$/WSe$_2$ heterobilayers  \cite{Seyler2019}; and tunable collective phases have been found to be present in twisted bilayer transition metal dichalcogenides \cite{Wang2020}.
These systems have emerged as interesting materials to  be  explored  under  twist  angles  and  moiré  patterns  since  they  lead  to  remarkable  effects  on  the electronic \cite{rosenberger2020twist,parzefall2021moire,Wang2020,ulstrup2020direct} and  optical  properties,  such  as  the  appearance  of  interlayer valley excitons  \cite{seyler2019signatures,parzefall2021moire,forg2021moire} and excitonic complexes such as the recently reported ``trions'' trapped in moiré superlattices \cite{marcellina2021,liu2021signatures}. These recent advances open new avenues to explore artificial excitonic crystals and quantum optic effects in these systems \cite{marcellina2021,liu2021signatures,guo2021moire}.

From a mathematical point of view, the underlying definition of the hexagonal commensurable cells presented some inconsistencies or unclear relationships with existing works via Diophantine Equation \cite{Shallcross2008,Shallcross2010a,hermann}.
A different general approach based on multilayer twisted affine lattices, known as Bravais-moir\'e theory, has been  applied independently to the existing reports on Diophantine equations and to the study of optical properties in square photonic crystals \cite{Gomez-Urrea2020a,Gomez-Urrea2020b}. It has also been applied to study graphene by introducing two unseen classes of twisted graphene with different electronic properties \cite{Tiutiunnyk2019}.

The aforementioned works have the common feature that the bilayer is projectively coincident with the reference lattice, i.e., when the points of the Bravais lattice of the second layer are projected onto the plane of the first layer they coincide exactly in space. However, there are materials which have the so-called displaced Bravais-moiré, with non-coincident projected layers, and therefore the problem of commensurate rotation must be considered.
This relatively unexplored situation in twistronic appears naturally in Chromium Trihalides bilayer compounds (CrX$_3$, X = Cl, Br, I), where one layer is shifted respect to the other. However the commensurable rotation in these types of systems has not been discussed so far. Furthermore, these magnetic compounds present unique opportunities to explore the moir\'e pattern effect on both their electronic and magnetic properties due to the strong interplay between magnetism and lattice 
degrees of freedom.
Among the CrX$_3$ family, the CrI$_3$ is considered a fascinating material that displays novel physical properties in the low dimensional limit. The magnetic interactions coupled to the lattice degrees of freedom give rise to FM or different AFM interactions depending on the number of layers and the structural phases. The understanding of these properties is being object of intense research recently with the focus on unraveling the interplay among magnetism, anisotropy, stacking and the layer limit (monolayers and bilayers)\cite{jiang2019stacking,VijayKumar,Lado2017,Leon2020}. Recent theoretical works have reported interesting approaches revealing that the magnetism is governed by mechanisms beyond super-exchange interactions \cite{Jang2019}, in which the interlayer exchange (FM/AFM) depends on the competition of the $d$-orbital interactions, which, in turn, depends on the different neighborhood generated by the stacking layers \cite{Sivadas2018,Jang2019}. These works suggest that this unique interplay between stacking and magnetism could have direct implications on moir\'e systems, and, in particular, in twisted CrI$_3$, where the stacking can dramatically alter the magnetic properties.

The innovation of twisted CrI$_3$ also involves a complex obstacle related to the metastability of the system when the shifted layers are rotated respect each other. The strong magnetic forces among the lattices do not guarantee that periodic stabilization can be achieved.
A very recent study in this direction evidenced the stability of moir\'e structures at low angles twist at around $1.0$-$1.36^{\circ}$ \cite{Xiao2021}. They exploit the  well-known property of approximated commensuration (quasi-commensurability) via very small rotation angles, which allows  certain long-period moiré patterns and useful mathematical simplifications for calculations. To the best of our knowledge, there are no works of exactly parametrized twisted magnetic bilayers with no use of approximated moiré periodicity. Therefore, we have focused on commensurable rotation angles in the CrI$_3$ bilayer in our research.

This work provides exact formulae for commensurable hexagonal super cells suitable for complex twisted systems as CrI$_3$. We explore formation of moir\'e structures at high rotation angles, namely 
$\theta$=$21.79^{\circ}$ and $32.20^{\circ}$, on CrI$_3$. First we discuss a general construction of moir\'e structure in non-shifted layers and then we present a general mathematical construction of moir\'e structures on shifted layers for the two well known stacking of CrI$_3$ the R$\bar{3}$ and the C2/m structural phases, known as low and high temperature structures, respectively. The main goal of this work is to show a general mathematical method for construction of moir\'e structures through shifted layers and to apply this methodology to explore the changes 
in the electronic, magnetic and structural properties of CrI$_3$ with moir\'e patterns. 

\section{Results and Discussion}
\label{sec:Model}

\subsection{Bravais-moir\'e Theory Under Shifted Twisted Lattices}
\label{subsec:BravaisMoire}

The Bravais-moir\'e theory deals with the problem of finding a super cell that reproduces $n$ projected layers on $R^2$ in a periodic way. The $n$ layers lie in parallel planes and are allowed to be different Bravais lattices and affine transformations of each other.  The classical literature considers  $n=2$  centered (AA-type stacking) or non-centered hexagonal (e.g. AB-type stacking) or square Bravais under identity transformations (trivial affine), which are similar and initially coincident, then a super unit cell is found through a commensurable rotation of the second layer as respect to the common origin with the first lattice. The so-called commensurable angle is obtained when a given position vector of the first lattice coincides with a position vector of the second lattice after performing the related rotation. Finding the commensurable rotation reduces to find the solution of a Diophantine equation based on a mapping of the integer pair $(r,s)$ (associated with a linear combination of the position vector in the first lattice) into an integer pair $(r',s')$ of the position vector after rotation of the second lattice.


Under the general affine approach, the commensurable angle $0<\theta<\pi/3$ for $n=2$ non centered hexagonal Bravais of the initially projected coincident lattices (stacking AA-type), is given by \cite{Tiutiunnyk2019}:

\begin{eqnarray}\label{eq1}
\cos \theta = \frac{r^2+6rs-3s^2}{6s^2+2r^2},
\end{eqnarray}  
where $r=2m+s$, with $m$ and $s$ two non-null natural coprime integers which satisfy $m>s$. Other equations, similar to Equation \ref{eq1}, have been reported to obtain commensurable angles with different parameterizations \cite{PhysRevB.47.15835, pong2005review, trambly2010localization}. These  different parameterizations must, however, give place to  the same solutions to the Diophantine equations that give rise to the commensurability. Therefore, all  the treatments of rotated bilayers in the AA stacking must match among them, except for their parameterization and the domain of validity of the parameters used.

\begin{figure}
\centering
{\includegraphics[width=\linewidth]{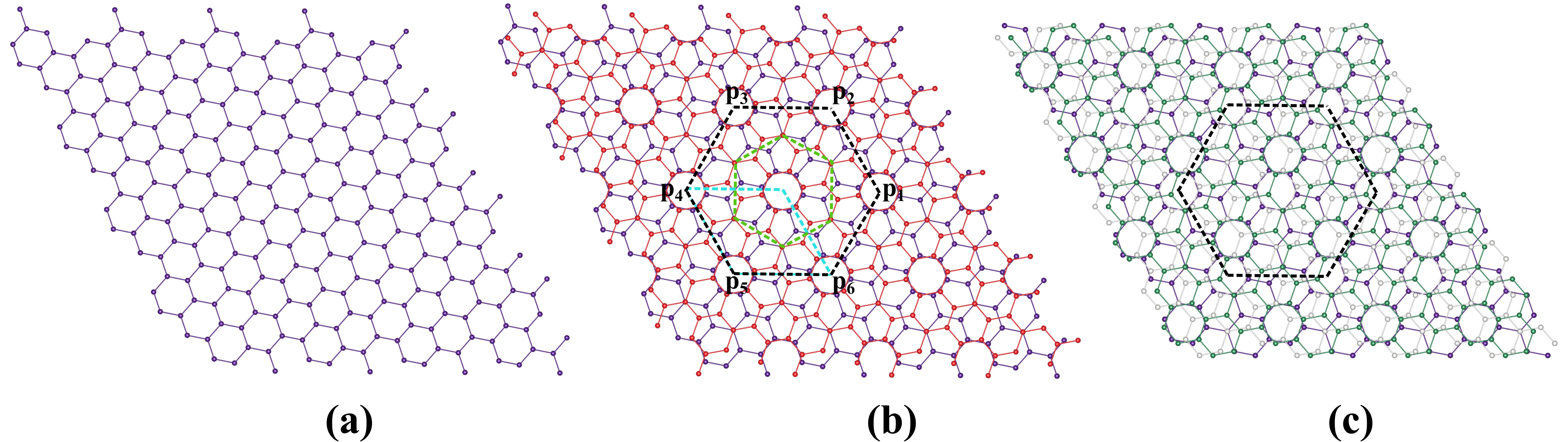}}
\caption{(a) Hexagonal cell L$_{fix}$ taken as a reference layer. (b) Bravais-moir\'e and super hexagonal cell built with $r$=5, $s$=1 by using Equation  (\ref{eq1}) . The commensurable rotation L$_{rot}$ (red dots) under $21.79^{\circ}$ is shown projected on the coincident reference system defined by $L_{fix}$ (purple dots). The primitive cell (cyan lines) and the affine points of the two layers, located at the vertices of the internal hexagon (light green dash lines) are also shown. (c) Shifted Bravais-moir\'e (green dots) indexed by $r$=5, $s$=1. This corresponds to a shifted cell from the hidden lattice which has a commensurable rotation L$_{rot}$ (gray dots) of $21.79^{\circ}$ with respect the reference system $L_{fix}$ (purple dots).}
\label{fig1}
\end{figure}

The rotation of the layer through this angle, L$_{rot}$, defines a commensurable hexagonal supercell located in the plane of the first layer, L$_{fix}$, with fixed Bravais lattice, which is taken as a reference, with vertices at the positions defined by:  

\begin{eqnarray}
p_{1} &=& k\left(\sqrt{3}r/2,3s/2\right)\\
p_{2} &=& k\left(-\sqrt{3}(3s-r)/4,3(s+r)/4\right) \nonumber\\
p_{3} &=& k\left(-\sqrt{3}(3s+r)/4,-3(s-r)/4\right)\nonumber\\
p_{4} &=&k\left(-\sqrt{3}r/2,-3s/2\right)\nonumber\\
p_{5} &=& k\left(\sqrt{3}(3s-r)/4,-3(s+r)/4\right)\nonumber\\
p_{6} &=& k\left(\sqrt{3}(3s+r)/4,3(s-r)/4\right),\nonumber
\label{vec}
\end{eqnarray}

where $k$ is the length of the reference unit lattice vector. \textbf{Figure \ref{fig1}(b)} shows the Bravais-moir\'e for $r=5$, $s=1$ with the corresponding commensurable rotation of $21.79^{\circ}$. Note the coincident points addressing the commensurable principle, as expected. 

Now, let us assume that an application involves two different Bravais that are not projectively coincident before the rotation (e.g AB-type stacking, AC-type stacking, etc.). Then we wonder for the commensurable rotation and the primitive vertices of the corresponding super cell. 
As far as we known, these types of commensurable systems are not reported so far, and the associated super lattice vectors and commensurable rotation need to be developed. The solution will be trivial if we adapt the principle given by the Equation (\ref{eq1}): \\

\textbf{Lemma 1.}
\textit{Let $H$ be a non centered hexagonal Bravais in $R^2$ spanned by the super lattice vectors  defined by the points  $p_{3}-p_{4}$ and $p_{5}-p_{4}$ under the commensuration angle $\theta$ in Equation (\ref{eq1}). Any new Bravais lattice $V$ constructed by a similarity (Euclidean) transformation of $H$ is also spanned by the same super lattice vectors  of $H$.\\}

For a proof, note that $V$ is just a scaling, rotation and/or translation of $H$; then $V$ can be seen as an atomic basis in the commensurable super cell defined by the lattice vectors of $H$. In other words, $V$ is just an Euclidean lattice spanning the same $R^2$, via the lattice original vectors of $H$.

We refer to the Bravais $H$ as the hidden or phantom lattice because it is used for generating the covering of $R^2$; meanwhile the resulting Bravais $V$ constructed by a rigid Euclidean transformation of $H$ will be named the visible lattice since it will keep the coordinates of the atoms of interest.

Finally, the so-called shifted-scaled or rotated Bravais-moir\'e can be easily constructed by commensurable rotation (Equation (\ref{eq1})). Let us assume that a fixed non centered hexagonal Bravais $L_{fix}$ is given. Let us construct a second visible Bravais $V$, which is a translation, rotation or a scaling of a hidden non centered hexagonal Bravais $H$ initially projectively coincident with $L_{fix}$.
According to the lemma, the visible Bravais $V$, which now can be refer to as $L_{rot}$ in the notation of Equation (\ref{eq1}), is a new Bravais-moir\'e just obtained by a commensurable rotation $\theta$ of $H$ respect to $L_{fix}$, and the hexagonal supercell vertices are the same $p_{1}\cdots p_{6}$ given for the classical twisted hexagonal bilayer. The resulting shifted-scaled or rotated Bravais-moir\'e hides the Bravais $H$, and only shows $L_{fix}$ and $L_{rot}$. 

The application of lemma 1 for $r=5$, $s=1$ is shown in \textbf{Figure \ref{fig1}(c)}. Here, the shifted Bravais-moir\'e lattice of the visible Bravais lattice, $V=L_{rot}$, represented by green points in the figure, is a rigid translation of the hidden lattice $H$ (gray points in the figure). 
$H$ is rotated $21.79^{\circ}$ respect to the fixed reference Bravais lattice $L_{fix}$, which is represented by purple points. The super cell is defined by the same vectors given by setting Equation (\ref{eq1}) for a classical Bravais-moir\'e indexed by $r=5$, $s=1$ and shown in \textbf{Figure \ref{fig1}(c)}, as well.

It is possible to apply this general construction to various 2D structures such as CrI$_{3}$, MoS$_{2}$, etc., by considering that a crystal structure consist of a Bravais lattice and a set of base atoms placed at each point of the lattice\cite{Kittel}. Thus, we explore the moir\'e structures for the CrI$_{3}$ with a non-shifted cell which has a symmetry group R3, and shifted cells with symmetry groups R$\bar{3}$ (low temperature ferromagnetic phase) and C2/m (high temperature antiferromagnetic phase).

\subsection{Structural and Magnetic Properties}

The formation of moir\'e patterns under commensurable rotations, following the method explained in previous section,   is applied to different stacking of the CrI$_3$ structure with R3, R$\bar3$ and C2/m symmetries. In each case, the top layer $L_{rot}$ is rotated respect to the fixed layer $L_{fix}$ with a commensurate angle of $21.79^{\circ}$  ($r$=5 and $s$=1) and $32.20^{\circ}$ ($r$=7 and $s$=1). These two angles correspond to the cells built with the smallest number of atoms under commensurable conditions. Table \ref{table-angN}  shows the amount of atoms needed in the primitive cell for simulating the systems at some selected twist angles. The smaller the commensurable twist angle the higher the amount of atoms needed, which strongly limits the possibility of performing DFT calculations.

\begin{figure}
\centering
{\includegraphics[width=\linewidth]{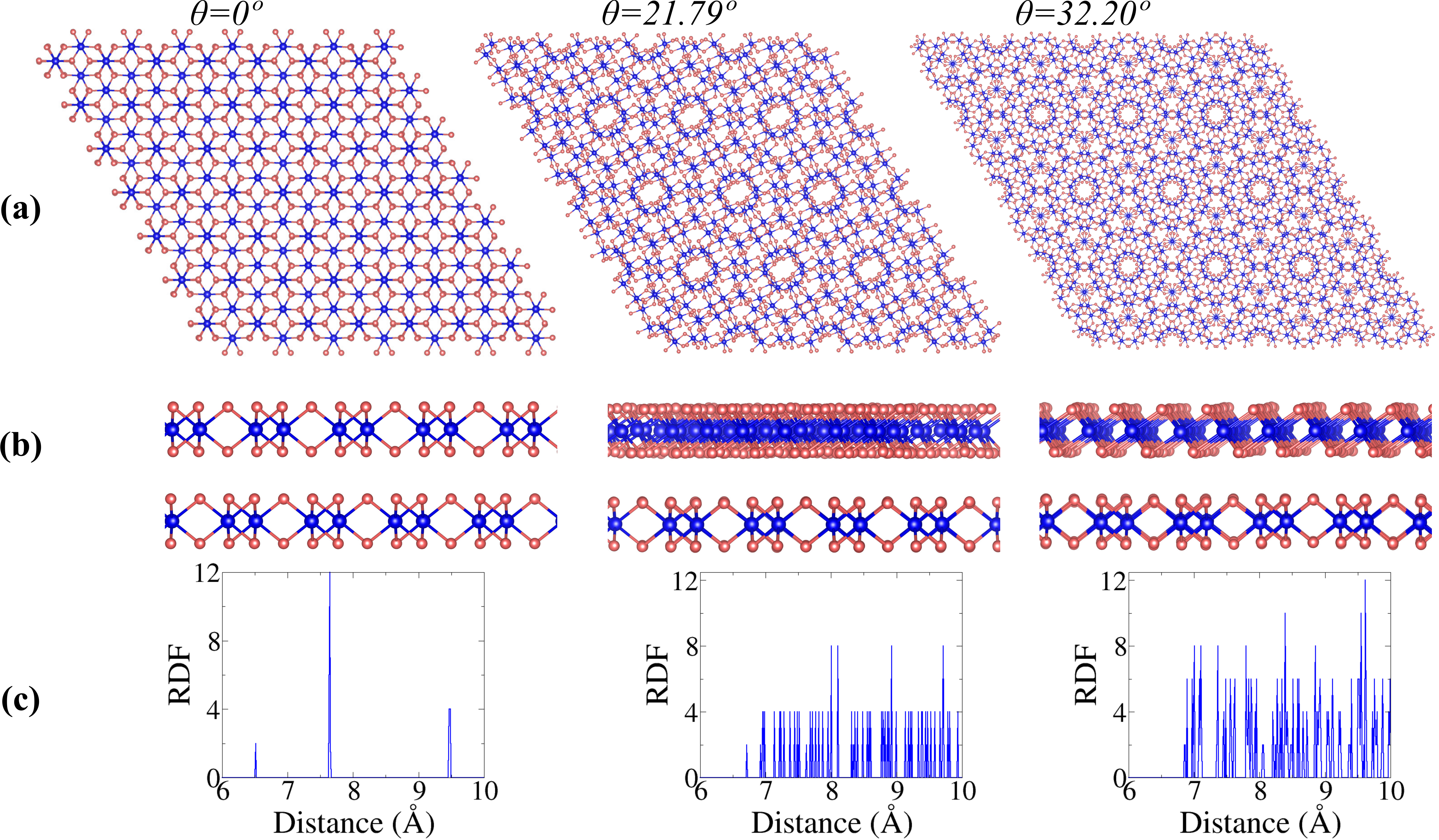}}
\caption{(a) Moir\'e patterns formed by relaxed structures for the studied angles. Blue balls stand for Cr atoms 
and light pink ones for I atoms. (b) Side view of corresponding structures (c) Radial distribution function (RDF) for interlayer Cr atoms. Only the lowest energy structure is reported for each $\theta$.}
\label{fig3}
\end{figure}

Table \ref{table-nn} reports the optimized lattice parameters, neighbors interlayer distances and magnetic order for each phase and for the rotation angles considered here ($0^{\circ}$, $21.79^{\circ}$  and $32.20^{\circ}$). 
\textbf{Figure \ref{fig3}(a)} shows the lowest energy Bravais-moir\'e crystalline structures, after relaxation, for each 
angle considered.




First, the lattice parameters are similar in each phase, when the rotation angle is fixed. For $\theta=0^{\circ}$, the R$\bar{3}$ structure has the highest cohesion energy, i.e., it is the most stable and can be considered as the fundamental state, as already reported previously \cite{jiang2019stacking,Jang2019,soriano2019,morell2019control}. The C2/m phase has an energy difference of 1.1 meV/atom with respect to the ground state, while the R3 phase is the structure with the lowest cohesion energy, with an energy difference of 4.7 meV/atom with respect to the ground state. Although this structure is a high energy metastable state, which has not been previously reported, it is important for the analysis of the rotated layers, as we will discuss
 below.

To describe the structural changes in each phase, we analyze the atomic coordination of Cr atoms by counting the number of nearest (n.n.) and next nearest neighbors (n.n.n) among Cr atoms belonging to different layers. We obtain the coordination interval to each atom from the radial distribution function (RDF) of each system, defining it as those atoms that lie around the peaks of the RDF, as shown in \textbf{Figure  \ref{fig3}(b}) and detailed in Table \ref{table-nn}. In this way we find that R$\bar{3}$ has the smallest interlayer distance with 1 and 9 n.n. and n.n.n., as reported \cite{Jang2019,Sivadas2018}. In C2/m and R3, the number of n.n. and n.n.n. change to (4,4) and (2,6), respectively, and they are at larger distances (see Table \ref{table-nn}). 

As a layer rotates over another one, there is a change in the distances among atoms and therefore there is a change in the neighborhood. As the tuning angle increases to $\theta=21.79^{\circ}$, the new configuration of R$\bar3$ has larger interlayer distances with respect to the phases with $\theta$=0 and with respect to the C2/m and R3 phases rotated by the same angle, as well. The number of n.n. and n.n.n. also changes with respect to 
$\theta=0^{\circ}$, giving now (10,6), (1,14), and (1,16) for the moir\'e systems R$\bar 3$, C2/m, and R3, respectively. Finally, by increasing the angle to $\theta=32.20^{\circ}$, the number of neighbors change to (3,17), (1,16), and 
(6,19), respectively.

These structural variations lead to important changes on the properties of these systems. Indeed, the most stable 
states change with respect to the non-rotated systems, with the lowest energy structure for the R3 phase on the two 
angles considered. The energy difference between the phases with higher and lower energy is 
$E_{R3}-E_{R\bar 3}$ = 0.3 meV/atom (for both $\theta=21.79^{\circ}$ and 32.20$^{\circ}$).

\begin{figure}
\centering
{\includegraphics[width=0.65\textwidth]{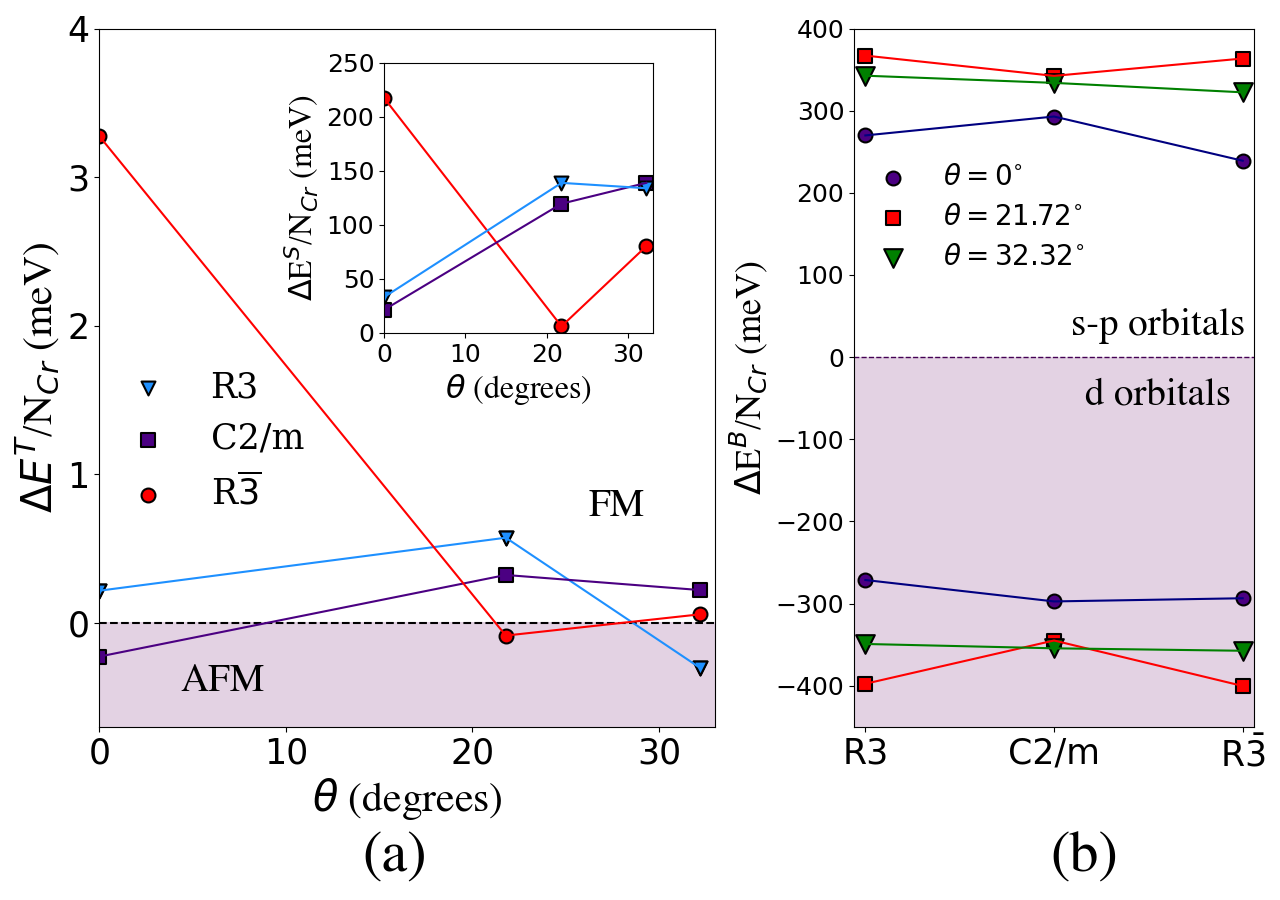}}
    \caption{(a) Total energy difference  $\Delta E^T$ per Cr atom ($\Delta E^T/N_{Cr}$, where $N_{Cr}$ stands for the number of Cr atoms) between AFM and FM configurations as function on the twist angle $\theta$.  
    (b) Contributions of $s+p$ and $d$ orbitals to the energy band $\Delta E^B$  per Cr atom ($\Delta E^B/N_{Cr}$) computed from the projected density of states on each orbital. The inset in (a) shows the energy difference, $\Delta E^S=\Delta E^T-\Delta E^B$, per Cr atom ($\Delta E^S/N_{Cr}$). The purple region represents an AFM ordering and the white one depicts the FM ordering.
  }
    \label{fig4}
\end{figure}

More important, due to the relative rotation between the two layers, transitions in the magnetic order in each of the phases are observed (see Table \ref{table-nn}). To explain theses changes, \textbf{Figure  \ref{fig4}(a)} depicts the energy difference between the FM and the AFM configurations ($\Delta E^T=E_{AFM}-E_{FM}$) as a function of the twist angle, for the three phases (the white and the colored regions represent the FM and the AFM states, respectively).


As a general trend, changes in magnetic order are observed as $\theta$ increases. At $\theta=0^{\circ}$, 
R$\bar3$ and R3 exhibit FM order with an energy difference of 3.45 and 0.417 meV/Cr-atom, respectively, while 
C2/m presents an AFM configuration with an energy difference of -0.23 meV/Cr-atom in agreement with previous reports\cite{Sivadas2018,Jang2019,soriano2019,Leon2020}.
By increasing the rotation angle to $\theta=21.79^{\circ}$, the systems R$\bar3$ and C2/m become AFM 
and FM, respectively, while R3 continues with the FM configuration. At $\theta=31.20^{\circ}$, R$\bar3$ and R3 go to an FM and AFM state, 
respectively, and C2/m continues with an FM configuration. The magnetic moments in each system are  $\mu_{Cr} = 3.4 \mu_B$ per Cr atom and $\mu_{I} = 0.1 \mu_B$, which are computed within spheres of Wigner-Seitz radii of 1.32 \AA \hspace{0.1cm}  and 1.49 \AA \hspace{0.1cm}
for Cr and I atoms, respectively. However, for the ferromagnetic state (at all the studied angles) the magnetic moment per Cr atom is obtained as 3.0 $\mu_B$, in perfect agreement with what is expected within the crystal field theory when considering Cr$^{+3}$ atoms in an octahedral field.
In summary, the following structural and magnetic changes are observed in the non-rotated states and the moir\'e systems:
\begin{eqnarray}
\nonumber E_{R\bar{3}} (FM) < E_{C2/m} (AFM) < E_{R3} (FM)   \hspace{0.2 cm} \theta =0^{\circ}~~~~~ & \\
\nonumber E_{R3}  (FM)  < E_{C2/m} (FM) < E_{R\bar{3}} (AFM)  \hspace{0.2 cm}  \theta =21.79^{\circ}  &\\
\nonumber E_{R3}  (AFM)  < E_{C2/m} (FM)  < E_{R\bar{3}}  (FM)   \hspace{0.2 cm}   \theta =32.20^{\circ} &
    \label{eq.1}
\end{eqnarray}

To explain the changes in the magnetic configuration relative to the non-rotated phases ($\theta=0^{\circ}$), we analyze the contributions to the total energy, which are mainly given by the long- and short-range interactions namely: 
$E^H$ (Hartree energy), $E^{II}$ (ion-ion interaction), $E^{XC}$ (exchange and correlation energy), 
E$^{vdW}$ (Van der Waals energy), and by the energy band $E^{B}$. The first contributions, 
$E^S= E^H+E^{II}+E^{XC}+E^{vdW}$, are directly related to the distribution in space of the ions and 
electrons composing the system; while $E^B$ is related to the quantum occupation of the electrons in 
the different energy levels and it governs the electronic structure of the system. Thus, the total energy can be written as $E^{T}=E^S+E^B$. In this framework, we can understand the FM and AFM order as a due to the competence between $E^B$ and $E^S$. According to our calculations, $E^S$ mainly contributes to stabilize the FM order, whereas $E^B$ contributes to stabilize the AFM configuration. In fact, the inset of \textbf{Figure  \ref{fig4}(a)} shows that, by removing the band energy, the original AFM systems became FM (see C2/m, R$\bar3$ and R3 in $\theta$ = $0^{\circ}$, $21.79^{\circ}$  and $32.20^{\circ}$, respectively), and the phases with FM order continue with the FM configuration. 
On the other hand, \textbf{Figure  \ref{fig4}(b)} shows that, in $\Delta E^B$, the $d$ orbitals of Cr atoms contribute to stabilize an AFM ordering and their contribution is greater than that of the $s-p$ orbitals of the I atoms, which tend to stabilize an FM ordering. 
Therefore, the following relations are fulfilled:

\begin{eqnarray}
\nonumber \Delta E^{S} &=& E^{S}_{AFM}-E^{S}_{FM} > 0 \\
\nonumber \Delta E^{B} &=& E^{B}_{AFM}-E^{B}_{FM} < 0.
\end{eqnarray}

In this framework, we can observe that the AFM state arises when the absolute value of the energy band change, 
$\mid \Delta E^B\mid$, is greater than the energy difference $\Delta E^S$ (i.e. $-\Delta E^B > \Delta E^S$). In the 
case of systems with FM state, the contribution of band energy does not suffice to compensate $E^S$.

The structural and magnetic changes from the FM R$\bar3$ phase with $\theta=0^{\circ}$ to the FM and 
AFM R3 phases with $\theta>0^{\circ}$ offer an optimum opportunity to show the interplay between 
magnetic interaction and crystalline stacking. To understand what determines this interplay, we start analyzing 
the change in stability for $\theta=0^{\circ}$. 
The R$\bar3$ phase has the lowest energy with respect to C2/m and R3 due to the smaller distance among  n.n. and n.n.n. of two magnetic atoms belonging to different layers (interlayer distance), as one can 
see in the Table \ref{table-nn}, $\theta=0^{\circ}$ column. Indeed, when the interlayer distance is smaller, the $E^S$ energy becomes more negative mainly due to an increase in the attractive Coulomb interaction respect to the repulsive one. This decrease in the interlayer distance also generates a larger difference $\Delta E^S$ between the FM and the 
AFM states, as it can be seen in the inset of \textbf{Figure  \ref{fig4}(a)}, resulting in a FM order, as mentioned above. 
This picture agrees with the suggested by previous work, where it is evidenced that the FM order of R$\bar3$ 
is due to the contribution to n.n.n, favoring the FM interactions \cite{Sivadas2018,Jang2019}. 
The same argument allows explaining why the R3 phase is more energetic and ferromagnetic for $\theta=0^{\circ}$.

For the case $\theta=21.79^{\circ}$, the R3 phase has a lower energy than R$\bar{3}$ due to the 
decrease of the n.n and n.n.n interlayer distances, while R3 has a lower energy than 
C2/m because, although the interlayer distances are equal, the number of n.n.n is 
larger (Table \ref{table-nn}). The FM order of R3 is due to the strong contribution of $E^S$. Or, from another 
viewpoint, due to the increase of the n.n.n over the n.n, favoring the  FM interactions \cite{Sivadas2018,Jang2019}. For $\theta=32.20^{\circ}$, R3 has still a lower energy 
because of the larger number of interlayer  n.n. n.n.  and n.n.n with respect to R$\bar{3}$, 
and because of the smaller distance and larger amount of n.n. than C2/m. 
For $\theta=32.20^{\circ}$, R3 has an AFM ordering because the increase in the interlayer n.n.  
distance contributes mostly with AFM interactions \cite{Sivadas2018,Jang2019} favoring a larger 
contribution of $E^B$ and thereby satisfying the condition $E^B<E^S$.

Although C2/m has always higher energy in both Bravais and Bravais-moir\'e, it is worth understanding the 
origin of their magnetic transitions from the AFM ($\theta=0^{\circ}$) to the FM configuration at 
$\theta=21.79^{\circ}$  and $32.20^{\circ}$ (\textbf{Figure  \ref{fig4}(a)}). These are either due to the decreasing distance between the interlayer n.n. and n.n.n regarding $\theta=0^{\circ}$, increasing $E^S$, or the n.n.n favoring the FM ground state \cite{Sivadas2018,Jang2019}. Moreover, a strong competition 
between the stability of C2/m and R3 (see binding energies in Table \ref{table-nn}) is observed, since both 
systems have similar neighbors (thus similar $E^T$). However, R3 is more stable because it has a larger amount 
of n.n. and n.n.n, allowing to increase the energy of the system, especially the contribution 
of $E^B$ as compared to C2/m (see \textbf{Figure \ref{fig4}(b)} at $\theta=21.79^{\circ}$  and $32.20^{\circ}$).
With the $\Delta E^T$ values  in Figure \ref{fig4}(a), the ordering temperature $T_{ord}$ can be obtained by using a Mean Field approximation \cite{Kittel}. For the  structures of minimum energy we obtain the $T_{ord}$ values 21.1, 3.7 and 1.95 K for $\theta = 0^{\circ}$, $21.79^{\circ}$ and  $32.20^{\circ}$, respectively. For the unrotated case, a value of 43.2 K was reported by  León \textit{et.al} \cite{Leon2020} using an Ising model, in agreement with our rough estimation of $T_{ord}$. We found a decreasing behavior of ordering temperature as the rotation angle increases.

Despite the magnetic effects on moiré superlattices have not been extensively explored, our results agree with previous studies indicating that changes in the neighbour and relaxation mechanisms play a key role in changing the electronic and structural properties of graphene  \cite{nam2017lattice, cantele2020structural, yananose2021chirality,lucignano2019crucial,koshino2019moire} and  metal dichalcogenides systems \cite{conte2019electronic,naik2020origin, mannai2021twistronics,angeli2021gamma}.

\subsection{Electronic Structure}

\begin{figure*}
\centering
\subfigure{\includegraphics[width=0.28\textwidth]{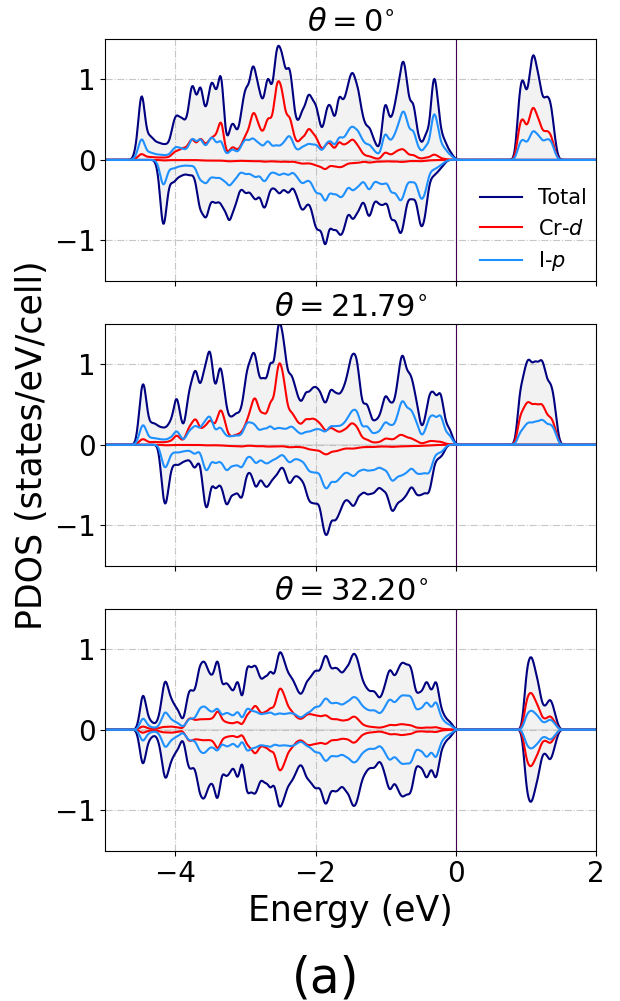}}
\subfigure{\includegraphics[width=0.28\textwidth]{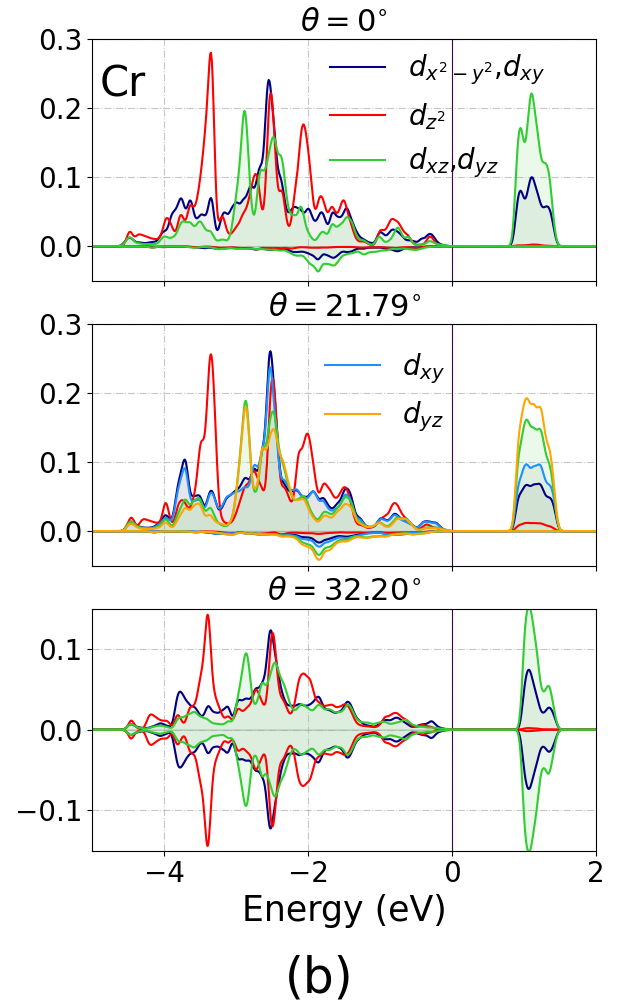}}
\subfigure{\includegraphics[width=0.28\textwidth]{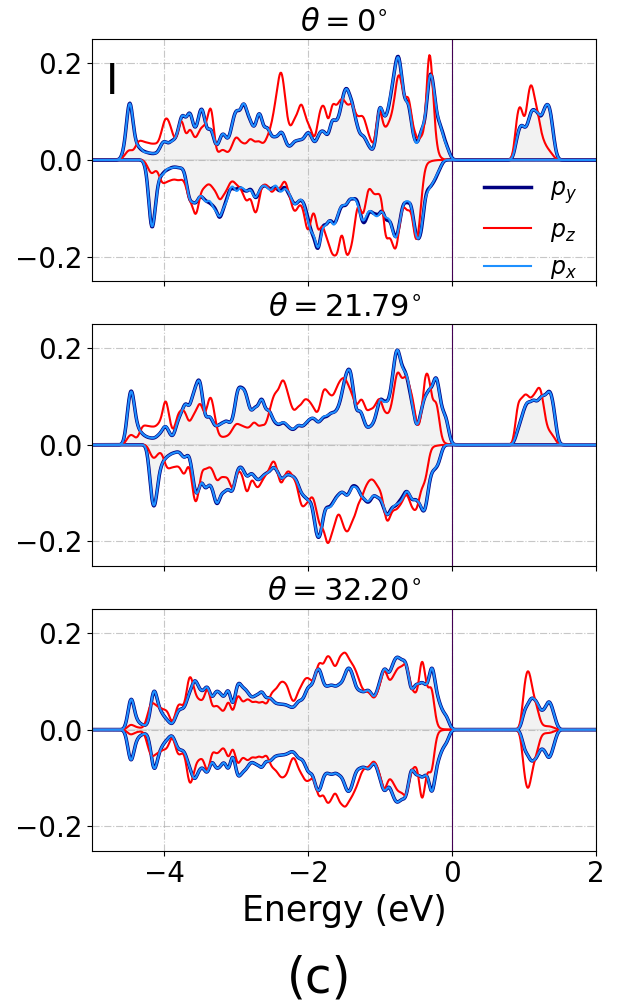}}
\caption{Total and projected density of states for the lowest energy phases (FM R$\bar{3}$ at $\theta=0^{\circ}$,  FM R3 at $\theta=21.72^{\circ}$ and  AFM R3 at $\theta=32.20^{\circ}$): (a) Total DOS, and  PDOS for Cr and I. (b) $d$-Cr PDOS, and (c) $p$-I PDOS.
}
\label{fig-dos}
\end{figure*}

\textbf{Figure \ref{fig-dos}} shows the total density of states (DOS) and the projected density of states (PDOS) for the minimum energy phases in each case, normalized by the number of atoms in the supercell used in these calculations. The projection of   DOS onto its different orbitals is done in the global coordinate system (with the coordinate axes set by the unit cell \cite{Kresse1996}). The considered systems behave as semiconductors with forbidden bandgaps of $E_g = 0.927$ eV, $E_g = 0.931$ eV and $E_g = 1.028$ eV for $\theta= 0^{\circ}$, $21.79^{\circ}$, and $32.20^{\circ}$, respectively. 
For all the phases, in general terms, it is observed that the valence and conduction bands are composed of 
$p$ orbitals of I and $d$ orbitals of Cr (\textbf{Figure  \ref{fig-dos}(a)}). In the valence band, there is a larger contribution 
from the $p$-I near the top of the band and a large contribution from the $d$-Cr in the middle of the band (between -3 and -2 eV, approximately).

For $\theta=0^{\circ}$, the $d$-Cr states are characterized by a strong contribution of $d_{z^2}$, in several 
energy intervals in the valence band, and by degenerate states: $d_{x^2-y^2}$ with $d_{xy}$ and 
$d_{xz}$ with $d_{yz}$ (\textbf{Figure  \ref{fig-dos}(b)}). In the conduction band there is no contribution from 
$d_{z^2}$ and it has the same degeneracy on the other states. On the other hand, the $p$-I states are 
characterized by $p_x$-$p_y$ degeneracy but with a similar contribution with $p_z$ (\textbf{Figure  \ref{fig-dos}(c)}). 
These results are in complete agreement with previous works \cite{VijayKumar,jiang2019stacking}.

For $\theta$=$21.79^{\circ}$, we observe changes in the occupancy states of the $d$-Cr states due to a 
break of the degeneracy $d_{x^2-y^2}$-$d_{xy}$ and $d_{xz}$-$d_{yz}$. The last one is due to reduction 
of symmetries when the second layer is rotated at this angle, as can be seen in the moir\'e patterns pictured  in \textbf{Figure \ref{fig3}(a)} (disorder of the atoms projected on the first layer is observed).  Moreover, there is a small 
contribution of $d_{z^2}$ in the conduction band.   The $p$-I orbitals do not undergo appreciable changes with respect to the non-rotated system. The degeneracy breakdown of the $d$-Cr orbitals negatively  increases the $E^B$ for the FM state but negatively decreases the AFM state due to the change of the 
interlayer distance, as discussed in the previous section.

The $\theta=32.20^{\circ}$ case is an AFM state and therefore the electronic population distribution of up and  down spins are different from that of the FM cases. For this angle, the states become degenerate again such 
as $\theta=0^{\circ}$, because the atoms of the second layer are spatially more ordered with respect to the  atoms of the first layer (see \textbf{Figure \ref{fig3}(a)}).

\begin{figure*}
\centering
\includegraphics[width=0.6\textwidth]{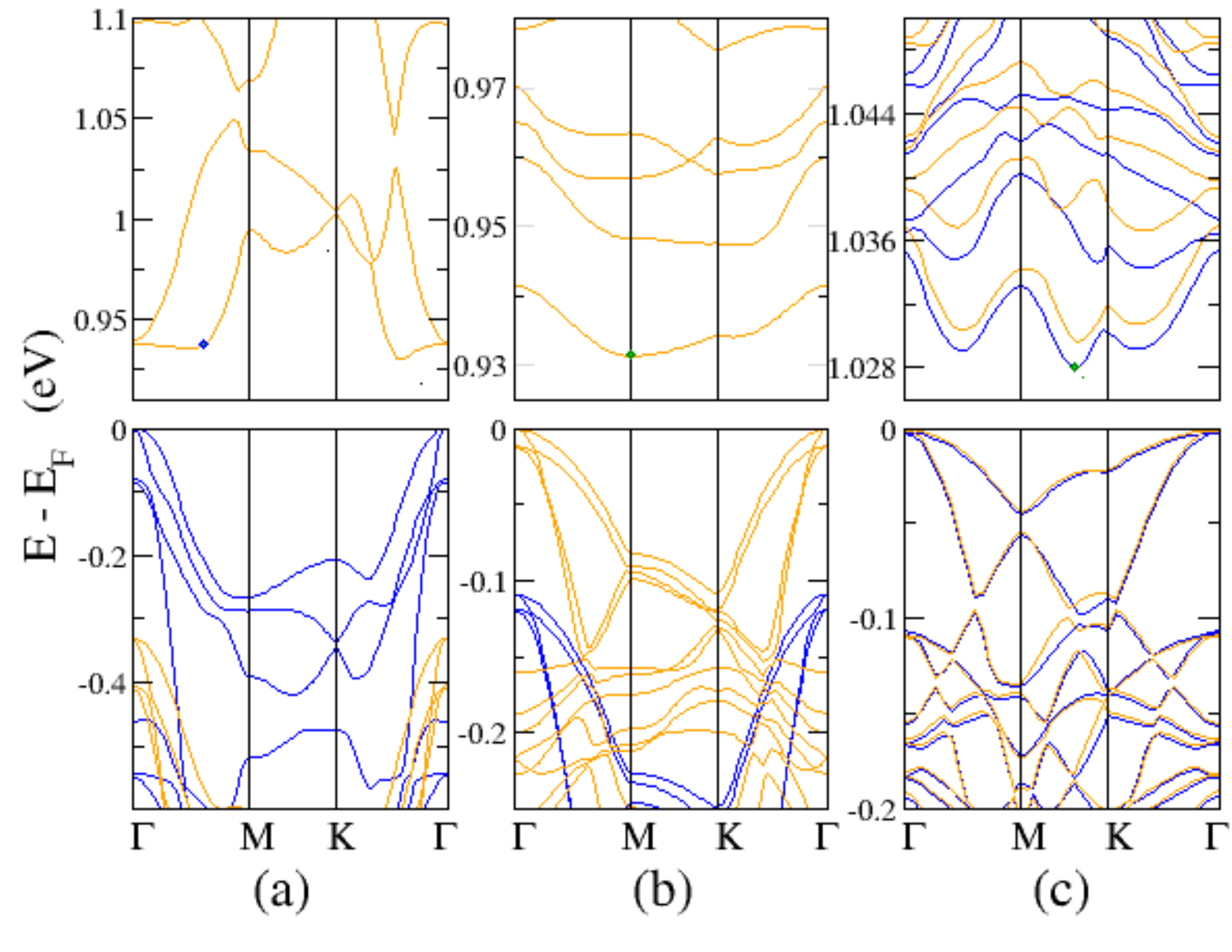}
\caption{Band structures for the lowest energy phases. (a) FM R$\bar{3}$ phase at $\theta=0^{\circ}$, (b) FM R3 phase at $\theta=21.79^{\circ}$ and (c) AFM R3 phase at $\theta=32.20^{\circ}$. A green circle shows the conduction band minimum (CBM) in each case. Blue (yellow) lines represent the spin up (down) channel. Moreover, the lower and upper panels correspond to the occupied and unoccupied bands, respectively.}
\label{fig-bands}
\end{figure*}

Band structures for the most stable phases are shown in \textbf{Figure \ref{fig-bands}}. It is observed that the valence 
band maximum (VBM) is at the $\Gamma$ point, while the conduction band minimum (CBM) changes 
when the second layer is rotated. For $\theta=0^{\circ}$ (\textbf{Figure  \ref{fig-bands}(a)}), VBM is mostly made up 
of $p_y$ states of I, and the CBM is in the middle of the K-$\Gamma$ zone with an energy difference of 
7.5 meV with respect to the $\Gamma$ point. This result agrees with calculations of Lado and Fern\'andez-Rossier\cite{Lado2017}, but differs from that by Gudelli and Guo\cite{VijayKumar} who report a direct gap. 
The VMB is mainly composed of $p$-I and $t_{2g}$-Cr orbitals, with higher contribution from the latter. 
The calculated effective masses for VMB and CBM are $-1.3 m_e$ and $2.4 m_e$, respectively.


For $\theta=21.79^{\circ}$ (\textbf{Figure  \ref{fig-bands}(b)}), CBM shifts to the M point and is also composed of $p$-I 
and $t_{2g}$-Cr orbitals as in the previous case. However, a break in the degeneracy of the states and 
changes in the occupancy of the different orbitals are observed, and VBM is now dominated by the $p_x$-I 
states and with less weight by the $p_y$-I states. In addition, a strong decrease in the band dispersion and 
changes in their curvatures are found, which increases the effective mass of the charge carriers. For this angle, 
there are two bands reaching the gamma point with effective mass of $-10.6 m_e$ for the heavy holes and 
$-1.8 m_e$ for the light holes. The effective mass of the electrons in the CMB increases to $70.7 m_e$, an 
order of magnitude larger than the non-rotated system.

For $\theta=32.20^{\circ}$ (\textbf{Figure  \ref{fig-bands}(c)}), CBM shifts again, this time toward the middle of the M-K zone, 
and it is also composed of $p$-I and $t_{2g}$-Cr orbitals as the previous cases. For this angle the break in the 
degeneracy of the states observed in $\theta=21.79^{\circ}$ disappears, and VBM is now mostly dominated by 
$p_x$-I states. Band dispersion still decreases and effective masses of $-19.5 m_e$ for the VBM and $15.5 m_e$ 
for the CBM are obtained. Since the magnetic state of this phase is AFM, \textbf{Figure \ref{fig-bands}(c)} shows the 
degeneracy of up and down spin bands in the valence band.

The new features found in these Bravais-moir\'e systems are the result of the strongest bonding originated 
from an enhancement of the direct interaction due to a hybridization increase of $p_{x,y}$ and $p_{z}$ 
orbitals (see \textbf{Figure  \ref{fig-dos}}). Similar characteristic have been found in moir\'e systems for low angles such 
as graphene \cite{Haddadi2020}, dicalgogenides systems \cite{naik2018,conte2019electronic}, few-layer graphite  \cite{ma2020}, carbon natubes \cite{arroyo2020}, among others. New studies on the electronic structure of Bravais-moir\'e systems varying 
the angle $\theta$ between $0^{\circ}$ and $360^{\circ}$ are necessary to have a clear trend regarding 
the electronic and magnetic tuning behavior.

\section{Conclusions}

We present a mathematical approach to perform commensurable rotations in shifted stacked systems for modeling moir\'e patterns. The method provides exact formulae for commensurable cells, without using approximations for long-period moir\'e patterns under small angles and large numbers of atoms.
From a mathematical point of view, the understanding of the algebra and geometry behind the resulting commensurable angles, after structural relaxation, opens up an interesting perspective for unknown non-Euclidean transformations and applications.

We have applied this method on CrI$_{3}$ bilayer systems to its two phases R$\bar{3}$, C2/m and to one 
non-shifted phase (R3 symmetry). Through DFT studies, we found that CrI$_{3}$ moiré structures with 
R$\bar{3}$ and C2/m stacking are metastable states, and the moiré non-shifted structure has lower energy at 
$\theta=21.79^{\circ}$ and $\theta=32.20^{\circ}$. These stability changes are directly related to the 
changes in the neighborhood due to the atomic arrangement that arises in the moiré structures. 
Furthermore, the emergence of moiré patterns on CrI$_{3}$ leads to important changes in its electronic 
properties with respect to the ground state. 

On one hand, we found magnetic phase transition from  
R$\bar{3}$-FM at $\theta=0^{\circ}$ to R3-FM and R3-AFM for $\theta=21.79^{\circ}$ and $32.20^{\circ}$, 
respectively. 
This result is compatible with a recent experimental work on this system \cite{xu2021}. The origin of these transitions are due to the changes in the distribution of the n.n. and n.n.n. In addition, we shown that the FM and AFM order can be understood as result of the competition between n.n. contributing to enhance the E$^{B}$, and satisfying the condition  $\Delta$E$^{B}$ $<$  $\Delta$E$^{S}$, favoring the AFM order. 
This framework agrees with results of previous works and offers a good opportunity to understand the 
AFM and FM order in moiré structures.

On the other hand, moiré structures in CrI$_{3}$ lead to changes in the band spectrum.  
The bigger the angle $\theta$, the less disperse the band. A increase of the electron mass in the valence band 
is obtained as a consequence of the increase in the hybridization bands between p$_{x,y}$ and p$_{z}$ orbitals of I atoms.
Our study suggests that the emergence of moiré patterns on CrI$_{3}$ could be a powerful tool for tuning 
magnetic and electronic states, opening new possibilities to explore electronic properties on CrI$_{3}$ and 
other related compounds.

\section{Computational Method}
\label{subsec:Computational}



Physical magnitudes such as total energies, magnetization, density of states, band structures, among others, 
are computed via First-Principles calculations by using DFT \cite{Hohenberg1964,Kohn1965} as implemented in the Vienna \textit{ab initio} simulation package (VASP) \cite{Kresse1996} with the PBEsol functional \cite{Perdew2008} and the optB86-vdW for the van der Waals correction \cite{vdw-1}. A Hubbard on-site Coulomb parameter (U=3 eV) is used for the Cr atoms to account for the strong electronic correlations, as suggested in a previous work \cite{jiang2019stacking}. For the structural optimization of the bilayers we include 20 \AA \hspace{0.1cm} of vacuum space to minimize the interaction between periodically repeated images along the $z$-axis. The structural optimization of the bilayers systems is performed within a force convergence of at least 10$^{-3}$ eV/\AA  \hspace{0.1cm} for each atom, and a plane-wave energy cutoff of 450 eV in order to relax the internal positions and the lattice parameters.  
Regular Monkhorst Pack grids of 10$\times$10$\times$1, 7$\times$7$\times$1 and 5$\times$5$\times$1, 
using supercells of 16, 112 and 208 atoms, are 
used to perform the ion relaxation for $\theta=0^{\circ}$, $21.79^{\circ}$  and $32.20^{\circ}$ cases, respectively. 
Increased grids of 12$\times$12$\times$1, 10$\times$10$\times$1 and 7$\times$7$\times$1 are used to 
perform the self-consistent calculations. 
In addition, to discard any possible dependence of our results on the DFT method, we performed the same calculations under the PBE functional with the van der Waals (vdW) approximation optB86-vdW \cite{vdw-1}  and  vdW-DFT-D2 \cite{grimme2006semiempirical}. Our results show the same trends for the electronic and structural properties, regardless of the method employed; which allows confidence in both the discussions and conclusions of our study.

.  The results show the same trends for the electronic and structural properties regardless the method employed.

\medskip
\textbf{Acknowledgements} \par
We acknowledge financial support from  ANID/CONICYT FONDECYT Grant Nº 1210193, ``Financiamiento basal para centros científicos y tecnológicos de excelencia AFB180001" and ANID/Becas Chile Postdoctorado Grant Nº 74190099. E. A. Velásquez and F. J. Caro-Lopera thank the University of Medellín for their support.

\medskip








%
\bibliography{iopart-num}


\newpage

\begin{table}
     \caption{Twisted commensurable angles following  the equation \ref{eq1} with the corresponding   amount total of atoms (N$_{Tot}$) into the supercell and amount of Cr atoms (N$_{Cr}$)  .}
   \centering
    \begin{tabular}{|c|c|c|c|c|}
    \hline
$\theta$ &	N$_{Tot}$&	N$_{Cr}$&	r&	s\\ \hline
5.1& 2032& 508& 20& 6\\ \hline
6.0& 1456& 364& 17& 5\\ \hline
7.3& 976& 244& 14& 4\\ \hline
9.4& 592& 148& 11& 3\\ \hline
11.0& 1744& 436& 19& 5\\ \hline
13.2& 304& 76& 8& 2\\ \hline
16.4& 784& 196& 13& 3\\ \hline
17.9& 1504& 376& 18& 4\\ \hline
21.8& 112& 28& 5& 1\\ \hline
26.0& 1264& 316& 17& 3\\ \hline
27.8& 640& 160& 12& 2\\ \hline
32.2& 208& 52& 7& 1\\ \hline
35.6& 1072& 268& 16& 2\\ \hline
38.2& 352& 88& 9& 1\\ \hline
42.1& 496& 124& 11& 1\\ \hline
44.8& 688& 172& 13& 1\\ \hline
46.8& 928& 232& 15& 1\\ \hline
  \end{tabular}
    \label{table-angN}
\end{table}

\begin{table}
     \caption{Optimized values for R3, R$\bar{3}$ and C2/m phases at $\theta=0^{\circ}$, $21.79^{\circ}$  and $32.20^{\circ}$. The following quantities are shown in order for each phase: the lattice parameter $a$=$b$ (\AA), the number of n.n.  and the distance interval (\AA) among  Cr atoms, the number of n.n.n.  and the distance interval (\AA) among  Cr atoms, the cohesive energy (eV), and the magnetic ordering.}
   \centering
    \begin{tabular}{c|l|c|c|c}
    \hline
    & & $\theta=0^{\circ}$ &  $\theta=21.79^{\circ}$  & $\theta=32.20^{\circ}$  \\  
 \hline
R3 &$a$ (\AA)&6.873 & 18.171 &  24.763 \\
 &n.n. [Interval (\AA)]&    2 [6.60]  &  1 [6.75] &  6 [6.85-6.89] \\    
 &n.n.n. [Interval (\AA)]&    6 [7.70-7.711] &  16 [6.92-7.29] & 19  [6.97-7.10] \\
  & Cohesive energy (eV)&  1.2150 &  1.2117 & 1.2118 \\
     &  Magnetic ordering&    FM & FM  & AFM \\
  \hline
   C2/m &$a$ (\AA)& 6.865 &  18.162& 24.762\\
  & n.n. [Interval (\AA)]&  4 [6.94-6.95]  &   1  [6.75]   &   1 [6.77].  \\
  & n.n.n. [Interval (\AA)] & 4 [7.99-8.02 ] &   14 [6.92-7.29] & 16 [6.86-7.10] \\
    & Cohesive energy (eV)&  1.2186 &  1.2116 & 1.2117 \\
       &Magnetic ordering&   AFM &  FM & FM \\
 \hline
 R$\bar{3}$ & $a$ (\AA)&  6.876 & 18.170  & 24.760  \\
   &n.n. [Interval (\AA)]& 1 [6.52] & 10 [6.94-7.07]  &3  [6.85]\\
   & n.n.n. [Interval (\AA)]& 9 [6.63-6.65]  &   6 [7.26-7.29] &    17 [6.94-7.10] \\
     & Cohesive energy (eV)&  1.2197 &  1.2114 & 1.2115 \\
       & Magnetic ordering&  FM &  AFM & FM\\
\hline
     \end{tabular}
    \label{table-nn}
\end{table}




\begin{figure}
\textbf{Table of Contents}\\

\medskip
\includegraphics[width=0.8\textwidth]{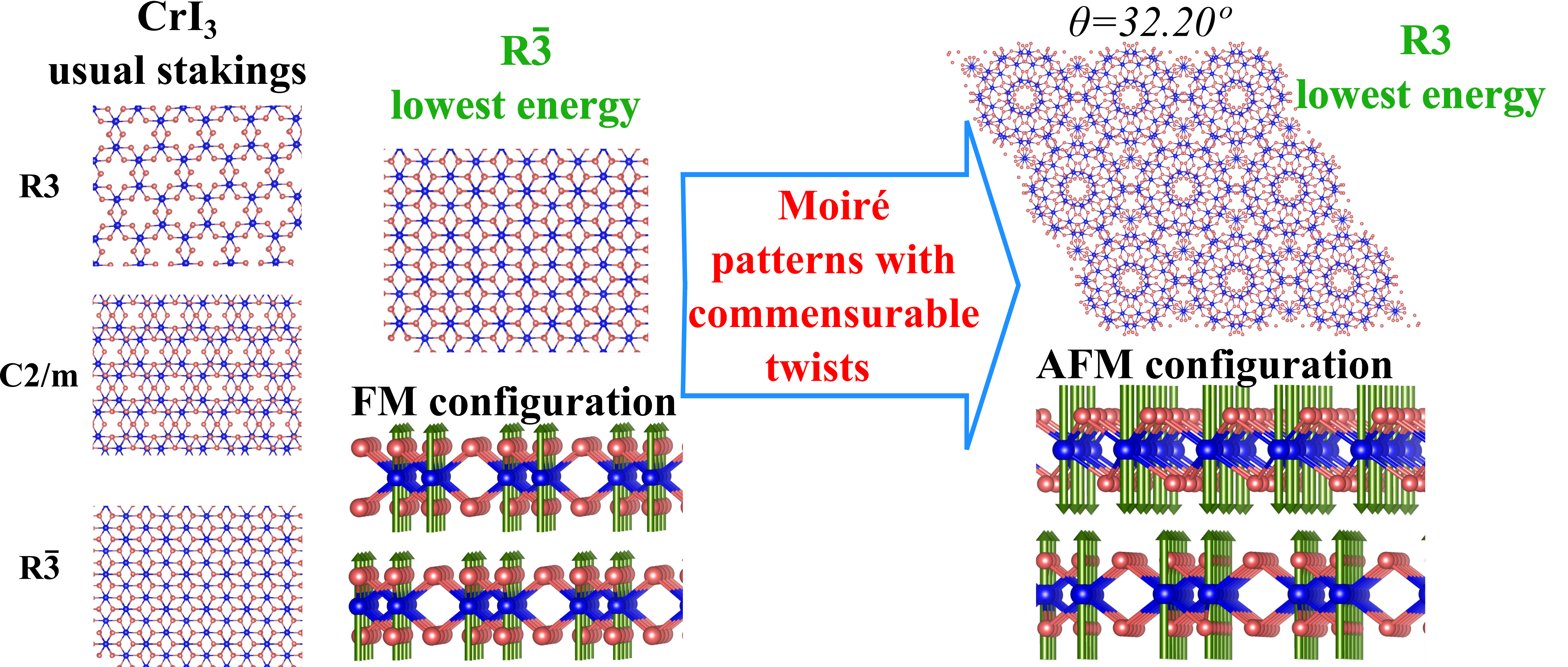}
  \medskip
  \caption*{moir\'e patterns with commensurable rotation on CrI$_{3}$ bilayer in its usual stacking phases R3, R$\bar{3}$ and C2/m are investigated. These commensurable rotations modify the magnetic ordering and the dispersion bands due to variations in the interlayer distance of nearest and second nearest neighbors of Cr atoms.}

\end{figure}

\end{document}